\documentclass[twocolumn,amssymb,nobibnotes,showpacs,superscriptaddress,aps,prl,longbibliography]{revtex4-1}

\usepackage{amsmath,graphicx}
\usepackage[colorlinks,linkcolor=blue,urlcolor=blue,anchorcolor=blue,citecolor=blue]{hyperref}

\begin{document}


\title{Dipole blockade without dipole-dipole interaction} 


\author{C. J. Zhu}
\affiliation{School of Physical Science and Technology, Soochow University, Suzhou 215006, China}
\affiliation{MOE Key Laboratory of Advanced Micro-Structured Materials,
	School of Physics Science and Engineering, Tongji University, Shanghai, China 200092}
\affiliation{Collaborative Innovation Center of Light Manipulations and Applications, Shandong Normal University, Jinan 250358, China}
\author{W. Li}
\affiliation{MOE Key Laboratory of Advanced Micro-Structured Materials,
	School of Physics Science and Engineering, Tongji University, Shanghai, China 200092}
\author{Y. P. Yang}
\email[Corresponding author:]{yang\_yaping@tongji.edu.cn}
\affiliation{MOE Key Laboratory of Advanced Micro-Structured Materials,
	School of Physics Science and Engineering, Tongji University, Shanghai, China 200092}
\author{G. S. Agarwal}
\email[Corresponding author:]{girish.agarwal@tamu.edu}
\affiliation{Institute for Quantum Science and Engineering, Department of Biological and Agricultural Engineering, Department of Physics and Astronomy, Texas A\&M University, College Station, Texas 77843, USA}


\date{\today}

\begin{abstract}
The dipole blockade phenomenon is a direct consequence of strong dipole-dipole interaction, where only single atom can be excited because the doubly excited state is shifted out of resonance. The corresponding two-body entanglement with non-zero concurrence induced by the dipole blockade effect is an important resource for quantum information processing. Here, we propose a novel physical mechanism for realizing dipole blockade without the dipole-dipole interaction, where two qubits coupled to a cavity, are driven by a coherent field. By suitably chosen placements of the qubits in the cavity and by adjusting the relative decay strengths of the qubits and cavity field, we kill many unwanted excitation pathways. This leads to dipole blockade. In addition, we show that these two qubits are strongly entangled over a broad regime of the system parameters. We show that a strong signature of this dipole blockade is the bunching property of the cavity photons which thus provides a possible measurement of the dipole blockade. We present dynamical features of the dipole blockade without dipole-dipole interaction. The proposal presented in this work can be realized not only in traditional cavity QED, but also in non-cavity topological photonics involving edge modes. 
\end{abstract}


\maketitle

Dipole blockade is a well-known effect in quantum optics, where only a single atom can be exited when the dipole-dipole interaction induced energy shift exceeds the linewidth of the excitation laser~\cite{saffman2010quantum}. If the excitation process is perfectly coherent, it is impossible to distinguish which atom is excited. Therefore, the system will be prepared in a symmetric superposition state that consists of all possible singly excited states. This superposition state leads to a very large concurrence in the dipole blockade regime~\cite{gillet2010tunable}. Moreover, the cooperative behavior induced by the dipole-dipole interaction will result in many extraordinary phenomena, such as entangled atoms~\cite{wilk2010entanglement,madjarov2020high,levine2018high,graham2019rydberg,saffman2009efficient}, enhanced Kerr nonlinearity~\cite{sinclair2019observation,bai2016enhanced,gorniaczyk2016enhancement}, superradiance behavior~\cite{dicke1954coherence,reimann2015cavity,gross1982superradiance}, single photon sources via the four wave mixing process~\cite{kolle2012four,brekke2008four,saffman2002creating}, blockade gate operation~\cite{tiarks2019photon,maller2015rydberg,muller2009mesoscopic}, collective encoding~\cite{pedersen2009few}, more complex quantum states of light and so on~\cite{nielsen2010deterministic,honer2011artificial,baur2014single,gorniaczyk2014single}.


The general physical mechanism for realizing the dipole blockade behavior is based on the large frequency shifts of the doubly excited state induced by the dipole dipole interaction~\cite{scully1999quantum,agarwal2012quantum}. Thus, the simultaneous excitation of two qubits is inhibited, and this effect is the more pronounced the stronger the dipole-dipole interaction is. In experiments, the distance between each two qubits is usually required to be smaller than the wavelength of the dipole transition to realize strong dipole blockade behavior. In past years, the dipole-dipole blockade effect has been observed in sample of Rydberg atoms due to its long range interaction~\cite{liebisch2007erratum,urban2009observation,gaetan2009observation,pohl2009breaking}. These early studies led to a lot of theoretical proposals for the quantum information processing and quantum computation~\cite{lukin2001dipole,browaeys2016experimental,saffman2005analysis}. Recently, theoretical and experimental efforts have been paid on the observation of the Rydberg blockade between two atoms~\cite{urban2009observation,gaetan2009observation}, and the realization of several atoms entanglement using a dressed Rydberg interaction~\cite{jau2016entangling,pupillo2010strongly,johnson2010interactions,sackett2000experimental}. 

In this letter, we propose a novel method to realize the dipole blockade  without dipole-dipole interactions and our proposal also yields a strong entanglement between two qubits. We consider a two qubits cavity QED system with a $\pi$ phase shift in positions of localized qubits as in experiments of Meschede~\cite{reimann2015cavity} and Rempe groups~\cite{neuzner2016interference}. Under the strong coupling regime, the symmetric Dicke state is decoupled with other states in the one-photon space. When the cavity decay rate is much larger than the emitting rates of qubits, the dipole blockade can be achieved just by adjusting the quality factor of the cavity. Contrary to the conventional dipole blockade where the double excited state is shifted far off resonance, we use the flexibility of using differently coupled qubits to the cavity and conditions on the relative decay of the qubit and the cavity mode. Moreover, it is possible to observe this cavity induced dipole blockade phenomenon by measuring the second-order photon correlation function. In the presence of this cavity induced dipole blockade, we also show that very strong entanglement between two qubits can be achieved. 


%
\begin{figure}[htb]
	\includegraphics[width=\linewidth]{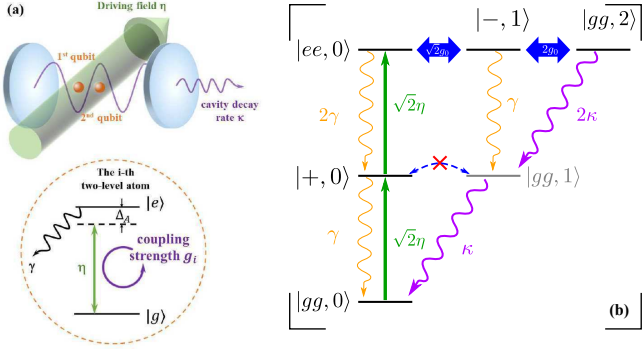}
	\caption{(a) Schematic diagram of a single mode cavity coupled with two two-level atoms driven by a pump field with angular frequency $\omega_P$ and driving strength $\eta$. The decay rate of each qubit is $\gamma$ and the cavity decay rate is $\kappa$. The coupling strength of the i-th qubit is $g_i$ ($i=1-2$), satisfying $g_1=-g_2$. (b) Physical picture leading to dipole blockade without dipole-dipole interaction. The transition pathways and qubit-cavity interactions for this two qubits QED system. The physical picture requires that the population of the doubly excited state should be quickly transferred to the cavity mode as we need to keep $\gamma$ very low so that the state $|+,0\rangle$ is significantly populated.}\label{fig:fig1}
\end{figure}
As shown in Fig.~\ref{fig:fig1}(a), we consider a scheme consisting of two identical qubits with resonant frequency $\omega_A$ in a single mode cavity, where the cavity resonant frequency is $\omega_{C}$, and two qubits are localized at different positions with a $\pi$ phase shift, yielding $g_1=-g_2=g_0$. A pump field drives these two qubits directly with angular frequency $\omega_P$ and driving strength $\eta$. Under the rotation wave approximation, the Hamiltonian of the system can be expressed as $H=H_0+H_I+H_L$ with  $H_0=\hbar\Delta_A(S^{(1)}_z+S^{(2)}_z)/2+\hbar\Delta_{C}a^\dagger a$, $H_I=\hbar g_0[(a^\dagger S_-^{(1)}+a S_+^{(1)})-(a^\dagger S_-^{(2)}+a S_+^{(2)})]$ and $H_L=\hbar\eta\sum_{i=1,2}(S_-^{(i)}+S_+^{(i)})$. Here, $\Delta_A=\omega_A-\omega_P$ and $\Delta_{C}=\omega_{C}-\omega_P$ are the detunings for the cavity and qubits, respectively. The operator $a^\dag$ ($a$) is the photon creation (annihilation) operator, while operators $S^{(i)}_z$ and $S_\pm^{(i)}$ are the spin operator of the $i$-th ($i=1,2$) qubit. 

In Fig.~\ref{fig:fig1}(b), we show the physical mechanism behind dipole blockade without dipole-dipole interaction. The figure \ref{fig:fig1}(b) is based on the use of collective states for two qubits. We illustrate the possible transitions, and the diagram also shows the condition needed to realize the dipole blockade. Using the collective states $|gg\rangle$,  $|\pm\rangle=(|eg\rangle\pm|ge\rangle)/\sqrt{2}$ and $|ee\rangle$ as basis, the Hamiltonian can be expressed in terms of the collective operators $D_z=S_z^{(1)}+S_z^{(2)}$ and $D^\dagger_\pm=(S_+^{(1)}\pm S_+^{(2)})/\sqrt{2}$, yielding $H_0=\hbar\Delta_AD_z/2+\hbar\Delta_{C}a^\dagger a$, $H_L=\sqrt{2}\hbar\eta(D_+^\dagger+D_+)$ and $H_I=2\hbar g_0(aD_-^\dagger+a^\dagger D_-)/\sqrt{2}$. Here, the symmetric and anti-symmetric Dicke states $|\pm\rangle$ are created by using the collective Dicke operators $D^\dag_\pm$~\cite{agarwal2012quantum,pleinert2017hyperradiance}. Assuming $\Delta_A=\Delta_{C}=0$, one can obtain a clear picture of  transition pathways for this two qubits cavity QED system. As shown in Fig.~\ref{fig:fig1}(b), the symmetric Dicke state $|+,0\rangle$ can be excited by absorbing a single photon (i.e., $|gg,0\rangle\overset{\sqrt{2}\eta}{\rightarrow}|+,0\rangle$). Due to the anti-symmetric coupling (i.e., $g_1=-g_2$), the symmetric Dicke state $|+,0\rangle$ is decoupled with other states in the one-photon space. By absorbing another photon, the state $|ee,0\rangle$ can be excited. In the two-photon space, possible qubit-cavity interactions is then via the coupling $|ee,0\rangle\overset{\sqrt{2}g_0}{\leftrightarrow}|-,1\rangle\overset{2g_0}{\leftrightarrow}|gg,2\rangle$. When the cavity decay rate is much larger than the emitting rates of qubits (i.e., $\kappa\gg\gamma$), the population of the doubly excited state should be quickly transferred to the cavity mode. Therefore, the probability of detecting the symmetric Dicke state $|+,0\rangle$ is predominant, yielding the occurrence of the dipole blockade induced by the cavity quality. It is worth to point out that this cavity induced dipole blockade phenomenon can also be detected by measuring the second-order photon correlation function $g^{(2)}(0)=\langle a^\dag a^\dag a a\rangle/(\langle a^\dag a\rangle)^2$. Since the photons leak from the cavity via the state $|gg,2\rangle$, strong bunching photons (i.e., $g^{(2)}(0)>2$) can be observed with the dipole blockade phenomenon.

We next present full master equation calculations to confirm the physical picture as outlined in Fig~\ref{fig:fig1}(b). To show this cavity induced dipole blockade quantitatively, one can directly solve the master equation (see the supplementary material) with damping terms of cavity and qubits, i.e., 
%
%
${\cal L}_\kappa\rho=\kappa(2a\rho a^\dagger-a^\dagger a\rho-\rho a^\dagger a)$ and ${\cal L}_\gamma\rho=\gamma\sum_{i=1,2}(2S_-^{(i)}\rho S_+^{(i)}-S_+^{(i)} S_-^{(i)}\rho-\rho S_+^{(i)} S_-^{(i)})$. The dipole blockade can be characterized by evaluating the ratio between the double excitation probability and the square of the single-excitation probability, i.e., $\xi\equiv P_{\rm ee}/P_{\rm e}^2$ under the steady-state condition. Here, $P_{\rm ee}=\langle ee|\rho_{\rm atom}|ee\rangle$ is the probability of finding two qubits excited, and $P_e=\langle e|\rho_{\rm atom1}|e\rangle=\langle e|\rho_{\rm atom2}|e\rangle$ denotes the probability of finding one of two qubits excited. In the presence of the dipole blockade, the probability of detecting the double excited state is smaller than that of detecting the single qubit excitation state. Therefore, the value of $\xi$ is smaller than unity, giving a direct signature of the dipole blockade effect. Stronger the dipole blockade effect becomes, smaller the ratio $\xi$ is.

\begin{figure*}[htb]
	\includegraphics[width=\linewidth]{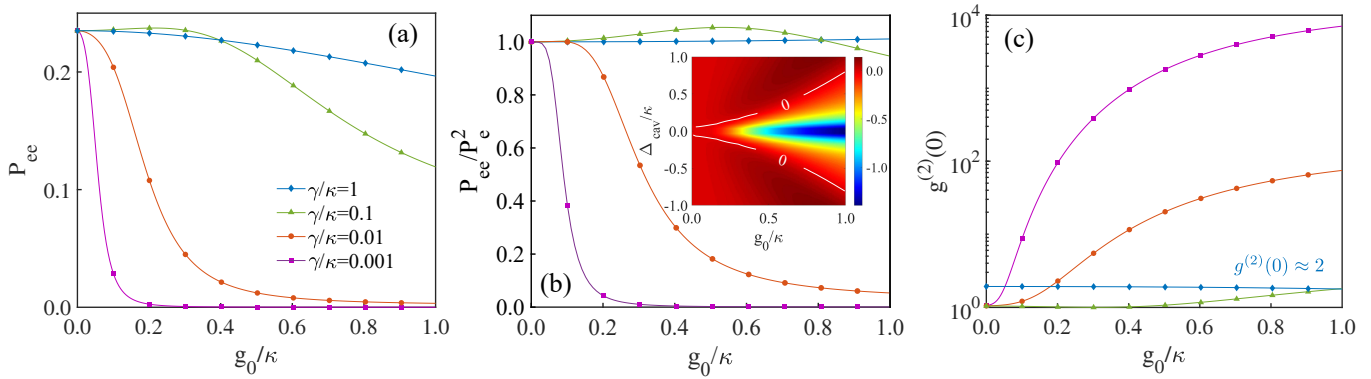}
	\caption{The probability of detecting the double excited state  $P_{\rm ee}$ [panel (a)], the ratio characterizing the dipole blockade $\xi$ [panel (b)] and the second order photon correlation function $g^{(2)}(0)$ [panel (c)] as a  possible method for detecting the dipole blockade in experiments. Here, the qubit decay rate is chosen as $\gamma/\kappa=1$ (blue), $0.1$ (green), $0.01$ (orange) and $0.001$ (purple), respectively. Other system parameters are given by $\Delta_A=\Delta_{C}=0$ and $\eta=5\gamma$. The inserted figure in panel (b) shows the value of $\log_{10}(\xi)$ as functions of the normalized coupling strength $g_0/\kappa$ and detuning $\Delta_{C}/\kappa$ with $\omega_{C}=\omega_A$. The white curves indicate $\log_{10}(\xi)=0$.}\label{fig:fig2}
\end{figure*}
Fist, we study the probability of detecting the double excited state $P_{\rm ee}$ as a function of the normalized coupling strength $g_0/\kappa$. The qubit decay rate is chosen as $\gamma/\kappa=1$ (blue curve), $0.1$ (green curve), $0.01$ (red curve) and $0.001$ (purple curve), respectively. Other system parameters are given by $\eta/\gamma=5$ and $\Delta_{A}=\Delta_{C}=0$. In the case of $\kappa\gg\gamma$, the value of $P_{\rm ee}$ [orange and purple curves shown in Fig.~\ref{fig:fig2}(a)] drops significantly as the qubit-cavity coupling strength increases. This is because the strong qubit-cavity interaction $|ee,0\rangle\overset{\sqrt{2}g_0}{\leftrightarrow}|-,1\rangle\overset{2g_0}{\leftrightarrow}|gg,2\rangle$ opens additional damping pathway and the decay of state $|gg,2\rangle$ becomes predominant. However, the symmetric Dicke state $|+,0\rangle$ uncouples from other states in the one-photon space due to the anti-symmetric coupling. Thus, it is stable and the dipole blockade occurs, yielding $\xi<1$ as shown in Fig.~\ref{fig:fig2}(b). In the case of $\kappa\leq\gamma$, the double excited state $|ee,0\rangle$ is as stable as the single excitation state $|+,0\rangle$. Thus, the dipole blockade phenomenon disappears and the ratio $\xi\approx1$ [see blue and green curves in panel (b)]. In the presence of the dipole blockade, the second order photon correlation function $g^{(2)}(0)>2$ shown in panel (c), and extremely strong bunching photons can be detected. 


In the inserted figure of panel (b), we also show the value of $\log_{10}(\xi)$ as functions of the normalized qubit-cavity coupling strength $g_0/\kappa$ and the detuning $\Delta_{C}/\kappa$, respectively. Here, we choose $\omega_{C}=\omega_{A}$, $\gamma/\kappa=0.01$ and $\eta=5\gamma$. It is noted that the cavity induced dipole blockade phenomenon only occurs in the resonant or near-resonant driving case. As the qubit-cavity coupling strength $g_0$ increasing, the frequency regime for realizing the dipole blockade is broadened simultaneously, which provides great facilities for the experimental implementation. 

\begin{figure}[htb]
	\includegraphics[width=\linewidth]{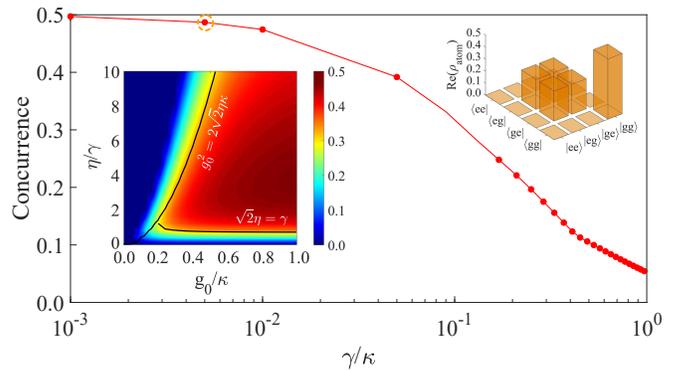}
	\caption{The red curve represents the steady state concurrence against the qubit decay rate $\gamma/\kappa$ with $g_0=\kappa$. The orange bar-diagram demonstrates the real part of the qubit density matrix elements with $\gamma=5\times10^{-3}\kappa$ (orange circle). The inserted diagram show the concurrence as functions of the normalized coupling strength $g_0/\kappa$ and driving intensity $\eta/\gamma$, respectively. Here, the black curves denote the boundary of the strongly entangled two qubits where the concurrence is close to $0.5$. }\label{fig:fig3}
\end{figure}
%
As demonstrated in Ref.~\cite{gillet2010tunable}, strong entanglement of two qubits can be achieved with the dipole blockade phenomenon. To show this interesting  property, we evaluate the concurrence of two qubits under the steady state condition, where the maximum value of the concurrence is $0.5$. As shown in the inserted diagram of Fig.~\ref{fig:fig3}, there exists a regime where these two qubits are strongly entangled (red area). The boundaries denoted by the black curves are given by $g_0^2=2\sqrt{2}\eta\kappa$ and $\sqrt{2}\eta=\gamma$, respectively (see the supplementary material for details). Here, the system parameters are chosen as $\Delta_A=\Delta_{C}=0$, $\gamma=0.01\kappa$. It is clear to see that the value of the concurrence tends to be saturated when the coupling strength $g_0$ is large enough.


Then, we choose $g_0=\kappa$ and scan the driving field intensity to show the influence of qubit decay rate to the entanglement of two qubits. In Fig.~\ref{fig:fig3}, the maximum value of the concurrence is calculated as a function of the normalized qubit decay rate $\gamma/\kappa$. Clearly, strong entanglement can be achieved when the qubit decay rate is much smaller than the cavity decay rate, e.g., the ratio $\gamma/\kappa\ll 0.01$. Except for the concurrence, the entanglement of two qubits can also be demonstrated by evaluating the real parts of the elements of the atomic density matrix $\rho_{\rm atom}$. When the ratio $\gamma/\kappa\gtrsim1$, the system is in the ground state $|gg\rangle$ in the equilibrium condition. However, in the case of $\gamma/\kappa\ll 1$, the system is in a mixture state consisting of a ground state with wight $0.5$ and a single qubit excitation state $(|eg\rangle+|ge\rangle)/\sqrt{2}$ as depicted by the orange bars with qubit decay rate $\gamma=10^{-3}\kappa$. 

Next, let's investigate how exclusive the condition of $g_1=-g_2$ to realize this cavity induced dipole blockade phenomenon and the corresponding two qubits entanglement. We first assume that one qubit is trapped at the peak of the cavity mode with a qubit cavity coupling strength $g_1=g_0$ for mathematical simplicity. Then, the coupling strength of the other qubit is given by $g_2=g_0\cos(\phi_z)$ where $\phi_z=2\pi\Delta_z/\lambda_{C}$ is the position dependent phase shift with $\Delta_z$ being the distance between two qubits and $\lambda_{C}$ being the wavelength of the cavity mode. 
\begin{figure}[htb]
	\includegraphics[width=\linewidth]{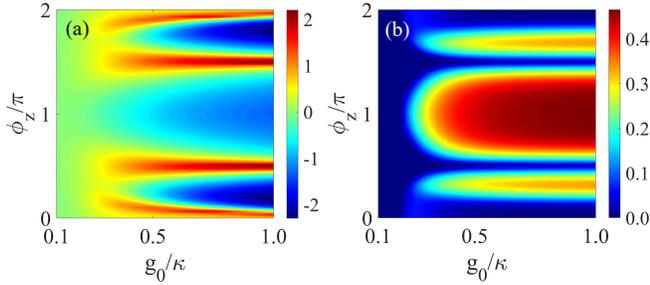}
	\caption{The variable $\log_{10}(\xi)$ [panel (a)] and the steady-state  concurrence [panel (b)] are plotted as functions of the normalized coupling strength $g_0/\kappa$ and the position dependent phase difference $\phi_z/\pi$ between two qubits. Here, we set $\gamma=0.01\kappa$, $\Delta_A=\Delta_{C}=0$ and $\eta=5\gamma$ for the numerical calculation.}\label{fig:fig4}
\end{figure}
In Fig.~\ref{fig:fig4}, we plot the ratio $\log_{10}(\xi)$ [panel (a)] and the concurrence [panel (b)] as functions of the normalized coupling strength $g_0/\kappa$ and the phase difference $\phi_z/\pi$, respectively. Here, we choose $\gamma=0.01\kappa$ and other system parameters are the same as those used in Fig.~\ref{fig:fig3}. It is noted that the cavity induced dipole blockade (i.e., $\xi<1$) and strong entanglement between two qubits (the concurrence is close to $0.5$) can be achieved over a wide regime of coupling strength near $\pi$ phase shift, which is robust for the experimental implementation. We also notice that the cavity induced dipole blockade can NOT be realized in the case of $g_1=g_2$ since the interaction between states  $|+,0\rangle$ and $|gg,1\rangle$ is allowed. More details is given in the supplementary material.  

\begin{figure}[htb]
	\includegraphics[width=\linewidth]{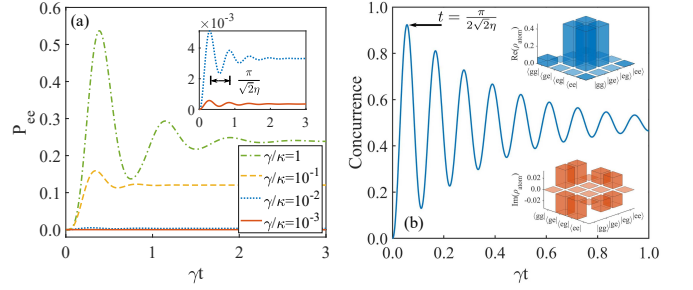}
	\caption{(a) The probability of detecting double excited state $P_{\rm ee}$ versus the normalized evolution time $\gamma t$ with the initial condition $P_{gg}=1$. The qubit decay rate is chosen as $\gamma=\kappa$ (green dash-dotted curve), $10^{-1}\kappa$ (orange dashed curve), $10^{-2}\kappa$ (blue dotted curve) and $10^{-3}\kappa$ (red solid curve), respectively. Other system parameters are given by  $\Delta_A=\Delta_{C}=0$, $g_0=\kappa$ and $\eta=5\gamma$. The inserted figure in panel (a) only shows the results of $\gamma/\kappa=10^{-2}$ (blue) and $10^{-3}$ (red), respectively. (b) The concurrence is plotted as a function of time with $\gamma/\kappa=10^{-3}$ and  $\eta=20\gamma$. The blue and red bars demonstrate the real and imaginary parts of the qubit density matrix at $t=\pi/(2\sqrt{2}\eta)$, respectively.}\label{fig:fig5}
\end{figure}

Finally, we discuss the dynamical properties of this cavity induced dipole blockade phenomenon and the corresponding entanglement of two qubits by numerically solving the time-dependent master equation with the initial condition $P_{gg}=1$. In Fig.~\ref{fig:fig5}(a), we show the probability of detecting double excited state $P_{\rm ee}$ as a function of the normalized time parameter $\gamma t$  with different qubit decay rate $\gamma=\kappa$ (green dash-dotted curve), $0.1\kappa$ (orange dashed curve), $0.01\kappa$ (blue dotted curve) and $0.001\kappa$ (red solid curve), respectively. Other system parameters are chosen as $g_0=\kappa$, $\eta=5\gamma$ and $\Delta_A=\Delta_{C}=0$. Obviously, the probability of finding double excited state reaches over $50\%$ in the case of $\gamma=\kappa$. As the qubit decay rate decreases, however, the double excited state drops quickly. In particular, the double qubits excitation will be significantly inhibited when $\gamma\ll\kappa$ (see blue and red curves in the inserted figure), resulting in strong dipole blockade phenomenon. The oscillation in $P_{\rm ee}$ can be explained by describing the system as a damping two-level model. Using the time-dependent Bloch equations, the probability of detecting the double excited state oscillates with a period of time $\pi/(\sqrt{2}\eta)$ (see supplementary material).

Likewise, the concurrence has similar dynamical characteristics exhibited in the profile of $P_{\rm ee}$. As shown in Fig.~\ref{fig:fig5}(b), the concurrence (blue solid curve) also oscillates with a period of time $\pi/(\sqrt{2}\eta)$ and reaches it maximum at time $t=\pi/(2\sqrt{2}\eta)$, corresponding to a maximum entanglement of two qubits. Here, we choose the system parameters as $\gamma=10^{-3}\kappa$, $g_0=4\kappa$ and $\eta=20\gamma$. The maximum value of the concurrence is close to $0.93$ at $t\approx0.06/\gamma$. The corresponding real [blue bars] and imaginary parts [red bars] of the elements of the atomic density matrix $\rho_{\rm atom}$ are demonstrated by the inserted figures in panel (b). Clearly, the probabilities of finding single qubit excited states are predominant compared with probabilities of finding qubits in other states. More importantly, these states are stable since their imaginary parts are zero. 

In conclusion, we provide a new physical mechanism for dipole blockade without the dipole-dipole interaction by using the cavity QED and putting two qubits at different positions. To realize the dipole blockade, the positions of two qubits are chosen suitably, and the decay strengths of the cavity field and qubits satisfy $\kappa\gg\gamma$. For example, in the case of a $\pi$ phase shift in positions of localized quits, the symmetric Dicke state is decoupled with other states in the one-photon space, and the population of the doubly excited state is transferred quickly to the cavity mode. Thus, the symmetric Dicke state is significantly populated, resulting in the dipole blockade behavior. Moreover, this effect can be detected by measuring the second-order photon correlation function. In the presence of the dipole blockade, extremely strong bunching photons can be observed. Using the mean field method, we also show strongly entangled two qubits can be achieved under the conditions of $g_0^2\geq2\sqrt{2}\eta\kappa$ and $\sqrt{2}\eta\gg\gamma$. With current experimental techniques, this cavity induced dipole blockade can be realized not only in traditional cavity QED systems, including the photonic crystal cavity with quantum dots~\cite{EnglundFaraon-2,VuQuantum2013}, the microwave cavity with Rydberg atoms~\cite{RaimondBrune-8} and circuit QED system~\cite{lehnert2003measurement,wallraff2004strong,blais2004cavity}, but also in non-cavity topological photonics involving edge modes~\cite{jiang2020seeing,guo2020sensitivity}, which makes the implementation of long range entanglement be possible.

\begin{acknowledgments}
CJZ thanks Prof. Hong Chen in tongji university for fruitful discussion. CJZ and YPY thank the support from the National Key Basic Research Special Foundation (Grant No.2016YFA0302800); the Shanghai Science and Technology Committee (Grants No.18JC1410900) and the National Nature Science Foundation (Grant No.11774262). GSA thanks the support of Air Force Office of  scientific Research (Award No. FA9550-20-1-0366).
\end{acknowledgments}

\bibliography{Refs}

\end{document}